\documentclass[twocolumn]{aastex62}
\graphicspath{{./}{figures/}}

\received{-}
\revised{-}
\accepted{26 March 2020}
\submitjournal{ApJ}

\shorttitle{MESAS: $\gamma$ Lep and $\gamma$ Vir}
\shortauthors{White et al.}

\begin{document}

\title{The MESAS Project: ALMA observations of the F-type stars $\gamma$ Lep, $\gamma$ Vir A, and $\gamma$ Vir B}

\correspondingauthor{Jacob Aaron White}
\email{jacob.white@csfk.mta.hu}

\author[0000-0001-8445-0444]{Jacob Aaron White}
  \affiliation{Konkoly Observatory,
  Research Centre for Astronomy and Earth Sciences,
  MTA Centre for Exellence,
  Konkoly-Thege Mikl\'os \'ut 15-17, 1121 Budapest, Hungary}

\author[0000-0001-9132-7196]{F. Tapia-V\'azquez}
\affiliation{Instituto de Radioastronom\'ia y Astrof\'isica, 
Universidad Nacional Autónoma de M\'exico, 
Morelia, 58190, M\'exico}

\author[0000-0002-3446-0289]{A.~G.~Hughes}
\affil{Department of Physics and Astronomy,
University of British Columbia,
6224 Agricultural Rd.,
Vancouver, BC V6T 1T7, Canada}

\author{A. Mo\'or}
  \affiliation{Konkoly Observatory,
  Research Centre for Astronomy and Earth Sciences,
  MTA Centre for Exellence,
  Konkoly-Thege Mikl\'os \'ut 15-17, 1121 Budapest, Hungary}
\affiliation{ELTE E\"otv\"os Lor\'and University, Institute of Physics, P\'azm\'any P\'eter s\'et\'any 1/A, 1117 Budapest, Hungary}

\author[0000-0003-3017-9577]{B. Matthews}
\affiliation{Herzberg Institute,
 National Research Council of Canada,
 5071 W. Saanich Road,
 Victoria, BC V9E 2E7, Canada}

\author[0000-0003-1526-7587]{D. Wilner}
\affiliation{Center for Astrophysics | Harvard \& Smithsonian,
60 Garden Street,
Cambridge, MA 02138, USA}

\author{J. Aufdenberg}
\affiliation{Physical Sciences Department,
Embry-Riddle Aeronautical University,
600 S Clyde Morris Blvd.,
Daytona Beach, FL 32114, USA}

\author{A. M. Hughes}
\affiliation{Department of Astronomy,
Van Vleck Observatory,
Wesleyan University,
96 Foss Hill Dr.,
Middletown, CT 06459, USA}

\author[0000-0003-0257-4158]{V. De la Luz}
\affiliation{Escuela Nacional de Estudios Superiores Unidad Morelia, 
Universidad Nacional Autónoma de M\'exico, 
Morelia, 58190, M\'exico}

\author[0000-0002-0574-4418]{A.~C.~Boley}
\affil{Department of Physics and Astronomy,
University of British Columbia,
6224 Agricultural Rd.,
Vancouver, BC V6T 1T7, Canada}



\begin{abstract}

The spectrum of stars in the submillimeter to centimeter wavelength range remains poorly constrained due to a lack of data for most spectral types. An accurate characterization of stellar emission in this regime is needed to test stellar atmosphere models, and is also essential for revealing emission associated with unresolved circumstellar debris. We present ALMA observations of the three nearby, main-sequence, debris-poor, F-type stars $\gamma$ Lep, $\gamma$ Vir A, and $\gamma$ Vir B at 0.87 and 1.29 millimeters. We use these data to constrain semi-empirical atmospheric models. We discuss the atmospheric structure of these stars, explore potential short term variability, and the potential impact on debris disk studies. These results are part of an ongoing campaign to obtain long wavelength observations of debris-poor stars, entitled Measuring the Emission of Stellar Atmospheres at Submillimeter/millimeter wavelengths (MESAS).

\end{abstract}



\section{Introduction} \label{sec:intro}

The dominant stellar emission mechanisms at submillimeter to centimeter (submm-cm) wavelengths depend strongly on the spectral type. The Sun, for example, has a corona with temperatures in excess of $10^{6}$ K coupled with many turbulent and variable processes that make it difficult to model at mm-cm wavelengths \citep[e.g.,][]{loukitcheva, wang, wedemeyer, delaluz14}. While these processes may be common for Solar-type stars, emission models for other spectral types have largely gone untested until only recently, due to the lack of observatories with the required sensitivity. In order to test stellar emission models at submm-cm wavelengths, observations of a broad range of spectral types, with no known circumstellar material, are required. The current sample of observed debris-poor main-sequence stars at long wavelengths is limited to only a few targets including $\alpha$ Centauri A/B \citep[G2V and K1V binary, Solar-like submm-mm spectra;][]{liseau15}, Sirius A \citep[A1V, much cooler photospheric emission in submm-cm;][]{white18, white19}, $\epsilon$ Eridani \citep[K2V, potentially separable long wavelength emission from the debris disk;][]{macgregor15}.

A broader and more precise understanding of stellar emission mechanisms is imperative for a better characterization of circumstellar environments in general. The smallest, coolest stars such as TRAPPIST-1 \citep[M8V star with 7 terrestrial planets][]{gillon17}, have been found to be frequent hosts of terrestrial planets at small orbital radii \citep[e.g.,][]{dressing15}. Due to strong stellar magnetic fields, high energy particles and radiation can impact the habitability which would otherwise nominally orbit within a habitable zone. Therefore, in order to accurately determine the habitability of its planets, TRAPPIST-1's submm-cm wavelength emission must be observed and modeled \citep{hughes19}. 

The clearing stages of planet formation will largely deplete the circumstellar environment of small $\mu$m - cm sized particles, as these are swept up into planets, asteroids, comets, or other large objects that are difficult or not possible to observe. As these larger objects collisionally evolve, they can replenish the small dust and debris in the system, creating what is referred to as a debris disk \citep[for a detailed overview of debris disk structure and formation, see][]{hughes18}. In an unresolved system, the presence of debris is typically inferred through modeling the expected stellar emission in the spectral energy distribution (SED) and assuming the excess is due to thermal emission from such a disk. Therefore, a reliable stellar model is required to detect and accurately study a debris disk. Debris disks are most commonly found around A-type stars \citep[e.g.,][]{su06,thureau}, and have an observed occurrence rate of $0.22^{+0.08}_{-0.07}$ around nearby FGK-type stars \citep{montesinos16}. With the exception of G-type stars, models of the submm-cm emission of these spectral types are largely non-existent. To properly constrain both the occurrence and abundance of excess emission due to circumstellar debris, we must first have well informed stellar spectra. 
 
In this paper, we present \textit{Atacama Large Millimeter Array} (ALMA) observation of $\gamma$\,Lep, $\gamma$\,Vir A, and $\gamma$\,Vir B. $\gamma$ Lep is a 1.3 Gyr main-sequence F6V star with a distant K2 companion at a separation of $\sim100''$ \citep{abt08, holmberg09}. It has no detectable amount of circumstellar debris, as evidenced by the lack of IR excess with \textit{Herschel}/PACS at $100~\mu$m and $160~\mu$m \citep[e.g.,][]{montesinos16}. $\gamma$ Vir is a binary system consisting of two 1.1 Gyr main-sequence F0IV stars with a $3\farcs6$ separation \citep{abt08, vican12}. \citet{montesinos16} noted that there was a ``dubious" detection of slight IR excess in \textit{Herschel}/PACS data which the nature and abundance are difficult to quantify due to the uncertainty in the photospheric fluxes and the stellar multiplicity. The similarities of the two stars in $\gamma$ Vir allows for testing of variability within a single observation if the separation is resolved. Together, these three stars are some of the closest F-type stars with no known debris, making them ideal targets for studying the stellar emission of F-type stars in the mm wavelength regime.

The details of the ALMA observations are presented in Sec.\,\ref{sec:obs}, the modeling is described in Sec.\,\ref{sec:model}, and the results and implications for debris disks are discussed in Sec.\,\ref{sec:disc}. These data are part of an ongoing effort to characterize stellar atmospheres at submm-cm wavelengths through Measuring the Emission of Stellar Atmospheres at Submillimeter/centimeter wavelengths (The MESAS Project). 

\section{Observations}\label{sec:obs}

\begin{table*}

\centering 
\begin{tabular}{c  c  c  c  c  c  c c} 
\hline\hline \ 
   	Star & Wavelength  & Date & Flux Calibrator & Flux  & Model Uncertainty & Obs. Uncertainty & Reduced $\chi^{2}$\\
   		& (mm)  & YYYY MMM DD & & ($\rm \mu Jy$)& ($\rm \mu Jy$) &  ($\rm \mu Jy~beam^{-1}$) & \\
   	\hline 
 
	$\gamma$ Lep   & 0.87 & 2018 Sep 22  & J0522-3627 & 707  & 17 & 30  & 2.83\\
	$\gamma$ Lep   & 1.29 & 2018 Sep 28  & J0522-3627 & 345  & 6  & 10  & 2.82\\
	$\gamma$ Vir A & 0.87 & 2018 Dec 20  & J1256-0547 & 653  & 13 & 22  & 3.52\\
	$\gamma$ Vir A & 0.87 & 2019 Mar 20  & J1256-0547 & 695  & 12 & 21  & 3.57\\
	$\gamma$ Vir A & 1.29 & 2019 Mar 09  & J1256-0547 & 323  & 8  & 14  & 3.49\\
	$\gamma$ Vir A & 1.29 & 2019 Mar 18  & J1256-0547 & 328  & 8  & 14  & 3.48\\
	$\gamma$ Vir B & 0.87 & 2018 Dec 20  & J1256-0547 & 681  & 13 & 22  & 3.51\\
	$\gamma$ Vir B & 0.87 & 2019 Mar 20  & J1256-0547 & 694  & 12 & 21  & 3.57\\
	$\gamma$ Vir B & 1.29 & 2019 Mar 09  & J1256-0547 & 307  & 8  & 14  & 3.49\\
	$\gamma$ Vir B & 1.29 & 2019 Mar 18  & J1256-0547 & 356  & 8  & 14  & 3.48\\

\hline	
    
\end{tabular}
\caption{Summary of observations and visibility model fitting results. The flux, model uncertainty, and reduced $\chi^{2}$ are from the {\scriptsize CASA} task \textit{uvmodelfit}. The observational uncertainty is the $\sigma_{rms}$ of each EB.  The stated uncertainties do not include the absolute flux calibration uncertainty, which is $\leq10\%$ at these wavelengths. 
}
\label{fit_par}
\end{table*}

We observed $\gamma$\,Lep and $\gamma$\,Vir with ALMA during Cycle 5 (ID 2017.1.00698.S, PI White) and Cycle 6 (ID 2018.1.01149.S, PI White). The observations of $\gamma$\,Lep used the J2000 coordinates of R.A. = $05^{\rm h}\, 44^{\rm m}\, 27^{\rm s}.79$ and $\delta =  -22^{\circ}\, 26' \,54\farcs18$ and the observations of the $\gamma$\,Vir system were centered on $\gamma$\,Vir A using the J2000 coordinates of  R.A. $= 12^{\rm h}\, 41^{\rm m}\,  39^{\rm s}.62$ and $\delta =  -01^{\circ}\,  26' \,57\farcs82$ with proper motion corrections at the time of each observation. The specific observational details for each target in each ALMA band are outlined in Secs. \ref{sec:lep} and \ref{sec:vir}.

We observed both targets with ALMA Band 6 and Band 7 and used instrument configurations with a total continuum bandwidth of 8 GHz split among 4 spectral windows (SPW). Each SPW has $128\times 15.625$ MHz channels for a total bandwidth of 2 GHz. The Band 6 SPWs were centered at 224 GHz, 226 GHz, 240 GHz, and 242 GHz, giving an effective continuum frequency of 233 GHz (1.29 mm). The Band 7 SPWs were centered at 336.5 GHz, 338.4 GHz, 348.5 GHz, and 350.5 GHz giving an effective continuum frequency of 343.5 GHz (0.87 mm).

All of the data were reduced using the Common Astronomy Software Applications ({\scriptsize CASA 5.4.1}) pipeline \citep{casa_reference}, which included water vapor radiometer (WVR) calibration; system temperature corrections; flux and bandpass calibration; and phase calibration. 

\subsection{$\gamma$\,Lep} \label{sec:lep}

The Band 7 observations were made on 2018 September 22 for 36.1 min (18.5 min on-source) and the Band 6 observations on 2018 September 28 for 65.3 min (42.0 min on-source). Both of the Execution Blocks (EB) used a 43 antenna configuration with baselines ranging from $15 - 1397$ m. The two EBs also used an identical calibration setup. Quasar J0522-3627 was the flux and bandpass calibrator; quasar J0609-1542 was the phase calibrator; quasars J0544-2241, J0609-1542, and J0522-3627 were used to calibrate the WVR. The average precipitable water vapor (PWV) was 0.24 mm for Band 7 and 0.97 mm for Band 6.

Both of the bands were imaged using the {\scriptsize CASA} \textit{CLEAN} algorithm using threshold of $\frac{1}{2}~\sigma_{rms}$ and natural weighting. The Band 7 data achieve a $\sigma_{rms}$ sensitivity of $30~\rm \mu Jy~beam^{-1}$ and the Band 6 data achieve a sensitivity of $10~\rm \mu Jy~beam^{-1}$ in the CLEANed images. The sizes of the resulting synthesized beams are $0\farcs25\times 0\farcs20$ at a position angle (PA) of $83.1^{\circ}$ for Band 7 and $0\farcs36\times 0\farcs30$ at a PA of $81.4^{\circ}$ for Band 6. These correspond to $\sim 2$ au and $\sim 3$ au at the system distance of 8.9 pc for Bands 7 and 6, respectively.

\subsection{$\gamma$\,Vir} \label{sec:vir}

The Band 7 observations were made on 2018 December 20 for 49.2 min (29.4 min on-source) and 2019 March 20 for 49.7 min (29.4 min on-source). The December EB used a 43 antenna configuration with baselines ranging from $15 - 500$ m and the March EB used a 43 antenna configuration with baselines ranging $15-313$ m. Quasar J1256-0547 was the flux and bandpass calibrator. Quasar J1229+0203 was the phase calibrator in December and quasar J1218-0119 was used in March. Quasars J1229+0203, J1256-0547, and J1222+0413 were used to calibrate the WVR in December, and J1218-0119, J1220+0203, and J1256-0547 were used in March. The average precipitable water vapor (PWV) was 0.70 mm in December and 0.40 mm in March. 

The EBs were imaged using the {\scriptsize CASA} \textit{CLEAN} algorithm using threshold of $\frac{1}{2}~\sigma_{rms}$ and natural weighting. Together, these Band 7 data achieve a $\sigma_{rms}$ sensitivity of $15~\rm \mu Jy~beam^{-1}$ in the CLEANed image. The size of the resulting synthesized beam is $0\farcs79 \times 0\farcs69$ at a position angle of $81.2^{\circ}$, corresponding to $\sim 8.7$ au at the system distance of 11.7 pc.

The Band 6 observations were made on 2019 March 09 for 52.8 min (34.5 min on-source) and 2019 March 18 for 52.9 min (34.4 min on-source). Both of the EBs used a 43 antenna configuration with baselines ranging from $15 - 312$ m. The two EBs also used an identical calibration setup. Quasar J1256-0547 was the flux and bandpass calibrator; quasar J1232-0224 was the phase calibrator; quasars J1218-0119, J1256-0547, and J1232-02243 were used to calibrate the WVR. The average precipitable water vapor (PWV) was 1.90 mm on March 09 and 1.82 mm on March 18.

The EBs were imaged using the {\scriptsize CASA} \textit{CLEAN} algorithm using threshold of $\frac{1}{2}~\sigma_{rms}$ and natural weighting. Together, these Band 6 data achieve a $\sigma_{rms}$ sensitivity of $10~\rm \mu Jy~beam^{-1}$ in the CLEANed image. The size of the resulting synthesized beam is $1\farcs43\times 1\farcs25$ at a position angle of $-65.2^{\circ}$, corresponding to $\sim 16$ au at the system distance.

\begin{figure*}
\centering
\includegraphics[width=\textwidth]{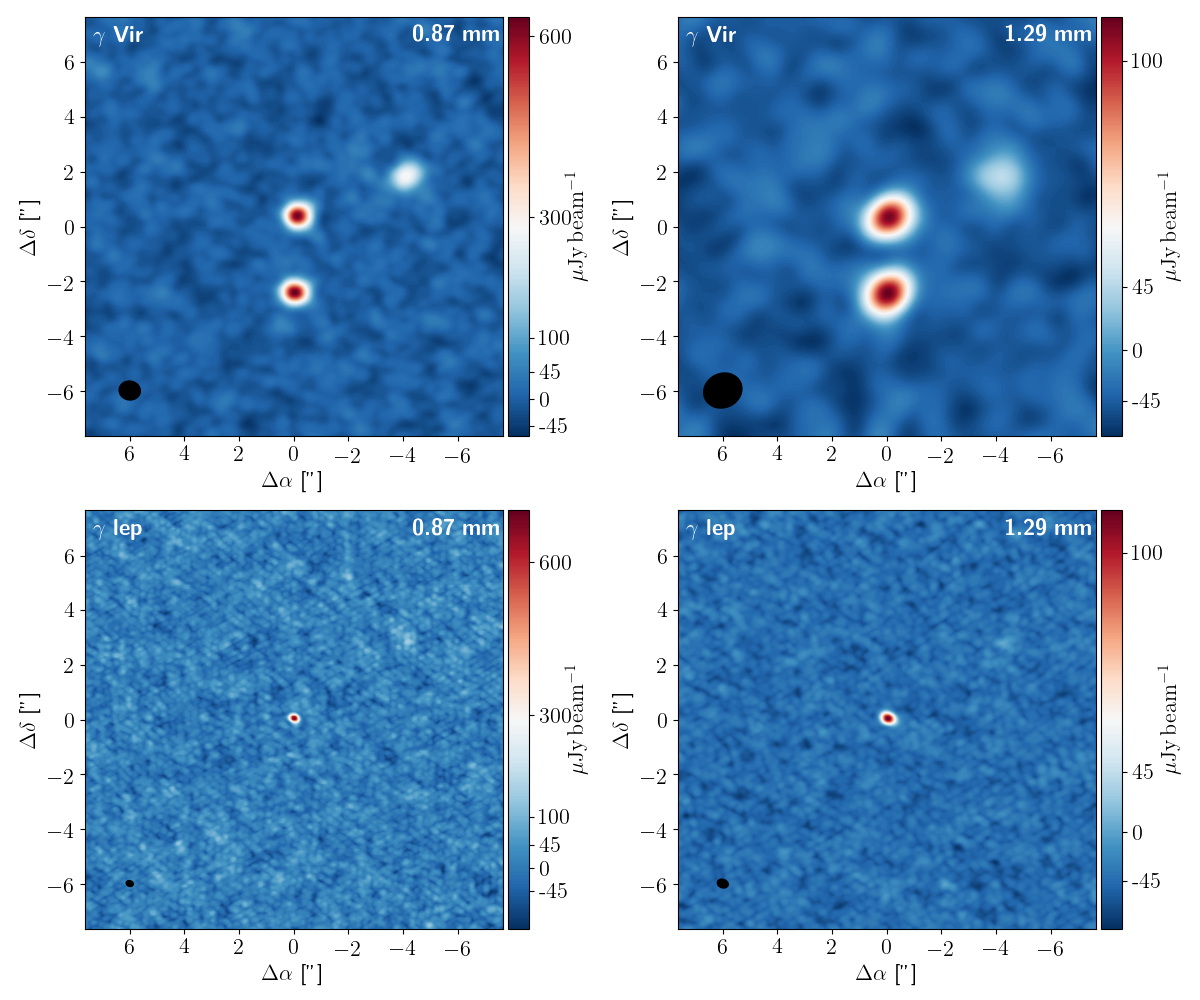} 
\caption{ALMA continuum observations of the $\gamma$ Vir system (top row) and the $\gamma$ Lep system (bottom row). For each system, the Band 7 (0.87 mm) observations are on the left and the Band 6 (1.29 mm) observations are on the right. $\gamma$ Vir A is located approximately in the phase center of the image, $\gamma$ Vir B is located to the South, and the unidentified background galaxy is located to the Northwest. The synthesized beam is denoted by the black ellipse in the bottom left of each image.  \label{vir_cont}
}
\end{figure*}

\section{Model Fitting}\label{sec:model}

\subsection{Visibility model fitting}\label{sec:uvmodel}

All of the stars observed are effectively point sources given the achieved synthesized beams. Therefore we utilize the same approach used in \citet{white18, white19} and obtain the flux using the {\scriptsize CASA} task \textit{uvmodelfit} and a point source model for all targets. This approach converges on a minimum $\chi^{2}$ through an iterative procedure. The results of the visibility model fitting are summarized in Table\,\ref{fit_par}. The flux uncertainties in Table\,\ref{fit_par} do not include the absolute flux calibration uncertainties, which we adopt as $\sim10\%$ (this is the commonly adopted uncertainty at these wavelengths per Sec.\,A.9.2 in the ALMA Proposer's Guide). As the same flux calibrator was used for both $\gamma$ Lep EBs, and the same for all $\gamma$ Vir EBs, the absolute flux uncertainty between observations of each system should be relatively minimal. 

We also explore the variability of each target, and within each EB, by using the same \textit{uvmodelfit} procedure applied to each on-source scan. We plot the time series for each star and discuss the results in Sec.\,\ref{sec:var}.

\subsection{Additional source near the $\gamma$ Vir system}

There is a third source present in the field of the $\gamma$ Vir system (see NW object in the top panels of Fig.\,\ref{vir_cont}). We were unable to identify the object in any catalogues, likely due to its very close proximity to $\gamma$ Vir (current projected separation of $\sim4\farcs2$). We detected the object at a signal-to-noise of $\gtrsim10$ in all EBs, so it is indeed a real source. The location of the peak flux is consistent between the 2018 December 20 EB and the 2018 March 08 EB to $<0\farcs05$, which implies that the object is likely not located within the foreground of the $\gamma$ Vir system (11.7 pc). The position angle of the object is different from that of the synthesized beam and the two $\gamma$ Vir stars, which implies that the source is at least marginally resolved. Given the galactic latitude of $+61^{\circ}$, it is unlikely that the object is a background circumstellar disk in a star forming region. 

To properly characterize the flux of this object, we fit both a point source and Gaussian model to the object with \textit{uvmodelfit} and find it to be more consistent with a Gaussian. The Band 7 total flux is $430\pm25~\mu \rm Jy$ and the Band 6 total flux is $160\pm16~\mu \rm Jy$.  This corresponds to a spectral index $\alpha\approx 2.5$ and a dust emissivity $\beta\approx 0.5$ assuming the object is in the Rayleigh-Jeans regime with $S_\nu \propto \nu^{2+\beta}$. These values do not appear to be consistent with the typical $\beta$ values of $1.5$ to $2.0$ \citep[e.g.,][]{casey14} for nearby dusty star forming galaxies. The angular size of $\sim1''$ and spectral index, however, can be more common for high redshift galaxies (R.\,Hill, priv. comm.). \citet{gonzalez-lopez20} use ALMA Band 6 number counts from the Hubble Ultra Deep Field to estimate the expected number of high redshift galaxies with a flux $>160~\mu$Jy to be $\sim19080~\rm deg^{-2}$, meaning the expected number of objects within our ALMA Band 6 field-of-view is about 0.5. Given all the considerations listed above, we conclude that the object is likely a background galaxy with either a high redshift or an anomalous spectral index. 

We use this object, along with the assumption that it has a relatively constant flux during each EB, to constrain the potential variability of the two stars in the $\gamma$ Vir system (see Sec.\,\ref{sec:var}).

\subsection{Stellar atmosphere models}\label{sec:atm_model}

\begin{table*}

\centering 
\begin{tabular}{c  c  c  c  c c } 
\hline\hline \ 
   	Parameter                   & Sun             & $\gamma$ Lep       &$\gamma$ Vir A & $\gamma$ Vir B \\
\hline 
	Distance, D (pc)                      & 0.0             & 8.93$^a$           & 11.68$^a$        & 11.68$^a$ \\
	Radius, R (R$_\odot$)                 & 1.0             & 1.32$\pm$0.04$^b$  & 1.45$^b$         & 1.45$^b$  \\
	Effective temperature, T$_{eff}$ (K)  & 5800$\pm$5$^c$  & 6255$\pm$70$^b$    & 6730$\pm$300$^d$ & 6694$^e$  \\
	Temperature minimum, T$_{R, min}$ (K) & 4400$^f$        & 4800           & 5580         & 5580  \\
	Minimum height, Z$_{min}$ (km)        & 560$^f$         & 750            & 625          & 625   \\	
\hline	   
\end{tabular}
\caption{Stellar parameters $^a$ \citet{gaia18}, $^b$ \citet{fuhrmann17}, $^c$\citet{munos13}, $^d$ \citet{kahraman16}, $^e$ \citet{boesgaard86}, $^f$ \citet{avrett}.}
\label{model_par}
\end{table*}

At far-infrared/mm wavelengths, stellar emission in FGK-type stars is dominated by optically thick free-free radiation \citep{dulk, gudel2002}. The flux is proportional to the plasma temperature (T$_R$) at a given wavelength and can be used to probe the temperature structure as a function of height above the photosphere. The stellar spectrum can therefore be used to build a model of the thermal structure of the chromosphere. In the Appendix (Figures \ref{GLP}, \ref{GVA}, and \ref{GVB}), we show the semi-empirical models for $\gamma$ Lep, $\gamma$ Vir A, and $\gamma$ Vir B. These models were generated using the KINICH-PAKAL code \citep{tapia20}. This code iteratively modifies the radial temperature and hydrogen density profiles, the ionization balance, and the opacity of a base model using the Levenberg-Marquardt algorithm to adjust the synthetic spectrum to the ALMA data presented here and the \textit{Herschel}/PACS data for $\gamma$ Lep \citep{montesinos16}. In the atmosphere, the chromosphere has a higher temperature than the photosphere leading to a strong deviation from radiative equilibrium. Therefore, we do not assume that ionization-excitation and radiative transfer are in local thermal equilibrium. For our models, a semi-empirical Solar model \citep[model C7 from][]{avrett} in hydrostatic equilibrium was adopted as the starting point. This can be taken as an average of the most commonly used Solar models \citep{vernazza81, fontenla93, loukitcheva}. For stars with a higher effective temperature than the Sun, this model serves as an initial condition. 
In order to constrain the region where the emission is generated in the models obtained from KINICH-PAKAL, we plot the contribution function (CF) \citep{tapia20} for each star at each wavelength. For $\gamma$ Lep, the Fig.\, \ref{GLP}(c) shows that the emission at millimeter wavelengths (1.29 mm and 0.87 mm) is generated at high chromosphere altitudes. At submillimeter wavelengths (0.16 mm and 0.10 mm) the emission is generated in the low chromosphere, close to photosphere. For $\gamma$ Vir A and $\gamma$ Vir B, Figs. \ref{GVA}(c) and \ref{GVB}(c) show that the radiation is generated around 2000 km over the photosphere.
The models for these three F-type stars were adjusted to observations made by ALMA and \textit{Herschel}/PACS, leading the synthethic spectrum to be in agreement with the observed one shown in Fig.\,\ref{tb_plot}. The brightness temperature for $\gamma$ Lep follows a Solar-like trend but $\gamma$ Vir A and $\gamma$ Vir B exhibit a large increase, likely due to the lack of constraining observations at longer wavelengths. The temperature structures and stellar spectra are discussed further in Sec.\,\ref{sec:disc_model}.

\section{Discussion}\label{sec:disc}

The ALMA data presented here are the first mm observations of F-type stars with no significant circumstellar material. These data allow for the first observationally constrained mm atmosphere models for stars of these spectral types, an assessment of the long wavelength variability/stability, and an exploration of the potential bias in interpreting unresolved circumstellar debris.

\subsection{Semi-empirical chromospheric model}\label{sec:disc_model}

\begin{figure}
\centering
\includegraphics[width=0.5\textwidth]{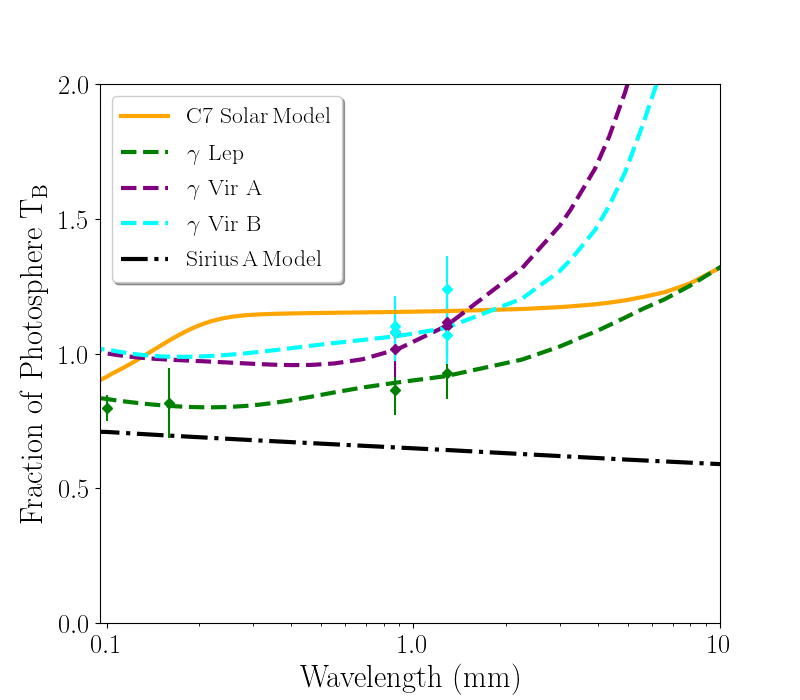} 
\caption{ ALMA observations of $\gamma$ Lep, $\gamma$ Vir A, and $\gamma$ Vir B. For illustrative purposes, we include the  orange curve, which is an average of the quiet and active Sun \citep[model C7 from][]{avrett} and the black curve, which is a PHOENIX model of Sirius A's atmosphere with a LTE model informed by previous submm-cm data as presented in \citet{white18, white19}. The green line and points represent the $\gamma$ Lep  model and observations. The purple and blue lines and points represent the models and observations of $\gamma$ Vir A and $\gamma$ Vir B, respectively. 
\label{tb_plot}
}
\end{figure}

For $\gamma$ Lep, the semi-empirical model is shown in Fig.\,\ref{GLP} as the green dashed lines. In the plasma temperature profile, we see a temperature minimum of T$_{R, min}$ = 4800 K (T$_{eff}$/T$_{R, min}$ = 0.76) at a height of 750 km above the photosphere. At this height, the model shows a hydrogen density of $n_H=9\times 10^{14}~ \rm cm^{-3}$. The high chromosphere has a positive temperature gradient until it reaches 2200 km where the transition zone begins.
The purple dashed line in Fig.\,\ref{GVA} show the semi-empirical model obtained for $\gamma$ Vir A. The plasma temperature profile presents two temperature drops one around 500 km and another close to 1750 km. In Fig.\,\ref{GVA} (c), the CF shows that radiation at 0.87 mm begins to form at 1530 km and 1600 km for 1.29 mm, with its maximum contribution for both around 2080 km.
For $\gamma$ Vir B, the semi-empirical model is shown as the blue dashed line in Fig.\,\ref{GVB}. This model has the same behavior as $\gamma$ Vir A in the plasma temperature profile. In Fig.\,\ref{GVB} (c), the CF shows that radiation at 0.87 mm  begins to form at 1600 km having its maximum contribution at 1970 km on the photosphere. For 1.29 mm, the greatest contribution is at 2120 km. In both models, the density is higher than the Solar average and the temperature profile shows that before 1500 km the plasma temperature is similar to the effective temperature.
For $\gamma$ Vir A and $\gamma$ Vir B, the plasma temperature anomalies below 1000 km and large brightness temperatures at long wavelengths (Fig.\,\ref{tb_plot}) are due to the fact that the KINICH-PAKAL framework uses observation data as constraints to model the temperature structure and the lack of data at shorter and longer wavelengths does not allow for restrictions of the plasma temperature profile in the upper and lower layers. Therefore, it is not possible to perform an analysis of the temperature minimum in these two stars.
The gray regions in the Fig.\,\ref{GLP}(a)-(c), Fig.\,\ref{GVA}(a)-(c), and Fig.\,\ref{GVB}(a)-(c) show the outer boundaries of the models where the convergence method cannot be applied directly.

The models presented here are a first step towards detailed models of the submm-cm spectra of F-type stars. The location and the depth of the temperature minimum in the chromosphere are influenced by the abundance of CO in the atmosphere \citep{linsky73} and acoustic waves \citep{Schmitz1981}. A combination of broad spectral coverage and CO spectra (e.g., with ALMA Band 9 and 10) will allow to investigate the formation height of T$_{R, min}$ and the temperature structure of the chromospheres.

\begin{figure*}
\centering
\includegraphics[width=0.49\textwidth]{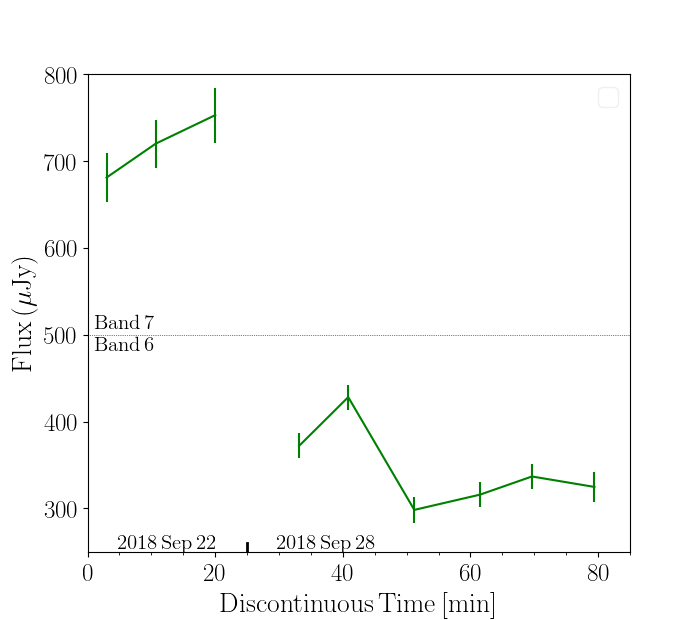} 
\includegraphics[width=0.49\textwidth]{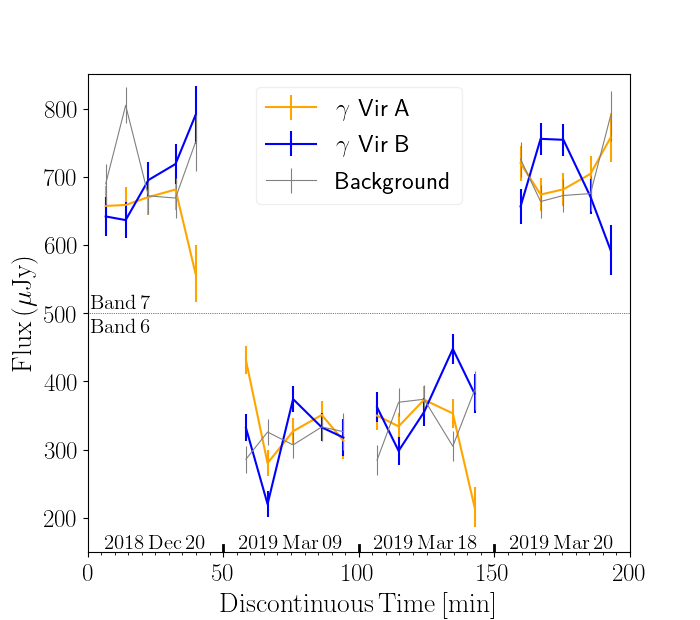} 
\caption{ Time series plots of $\gamma$ Lep (Left) and the $\gamma$ Vir system (right). For each point on the plot, the total flux was fit using the {\scriptsize CASA} task \textit{uvmodelfit} and broken up into $6-8$ minute time intervals. The date of each EB, and corresponding times series, is denoted on the bottom of each plot. While the time is plotted in minutes, we note that it is not continuous between EBs. The relative time between data points, however, is preserved. For $\gamma$ Vir, the flux of the background source has been shifted up to coincide with the two stars. This approach was used as the EBs were observed weeks to months apart and only a qualitative assessment of the variability within a single EB is being considered. \label{time_plots}
}
\end{figure*}

\subsection{Potential millimeter wavelength variability}\label{sec:var}

The stellar atmosphere models presented in Sec.\,\ref{sec:atm_model} reproduced the observed flux reasonably well if the observations are combined over the entire time on-source in each EB to increase the signal-to-noise. As is common practice in radio interferometry, the EB observing strategy alternates between the target and a phase calibrator until the total on-source time is achieved. We can therefore further divide up the observations into increments of $6-8$ minutes between individual phase calibrations to assess potential short term variability. For $\gamma$ Lep, we show the time series flux for both Bands 6 and 7 on the left side of Fig.\,\ref{time_plots}. While the curves are indeed not flat, they are roughly consistent with the full integrated flux and there is likely no variability at the level of the flux calibrations over the course of the observations. As these are the first mm wavelength observations of $\gamma$ Lep, we cannot rule out the possibility of variability at time periods longer than the length of the EBs.

For the $\gamma$ Vir system, we have the benefit of observing a binary with two nearly identical stars at multiple dates, as well as a background source (with no presumed variability). The times series is shown on the right side of Fig.\,\ref{time_plots}. For random or uncharacterized observational uncertainties, we would expect both $\gamma$ Vir A and $\gamma$ Vir B to follow a similar trend, as they are both located near the phase center of the observations (i.e., the phase calibration should be better towards the phase center). The background source is included too, as it is unlikely to exhibit any significant variability if it is indeed a high redshift galaxy (we shifted the flux of this source up to match the $\gamma$ Vir stars for ease of presentation). Therefore, we can reasonably assume that the apparent variability in the background source is consistent with observational uncertainty on such short time intervals. We plot the time in minutes but note that it is discontinuous between EBs. Plotting the times series with this approach is the best way to compare all 4 of the EBs which were observed at irregular intervals over a 3 month period. The spacing between data points within a single EB is preserved, allowing for general, short period trends to be identified.

When comparing the 3 objects in the $\gamma$ Vir field, we find that $\gamma$ Vir A and $\gamma$ Vir B at times deviate from the trend of the background source and each other. For example, the Band 7 observations from 2019 March 20 show that the background source and $\gamma$ Vir A have a nearly identical trend while $\gamma$ Vir B has a markedly different trend. In contrast to this, during the second half of the 2018 December 20 observations the background source and $\gamma$ Vir B have a similar trend while $\gamma$ Vir A is the star that appears to deviate. Taken at face value, these time series data suggest that there is potentially variability in one or both of the $\gamma$ Vir stars at the minute to 10s of minutes time scales. We caution though that there could still be some unconstrained systematic uncertainty in each flux value that is driving the observed trends. Future observations are necessary in order to confirm or reject the presence of millimeter variability in the $\gamma$ Vir system.

Short period variability that is difficult to model is not completely unheard of for stars. \citet{balona1994} found $\gamma$ Doradus to be of a new class of short period pulsating F-type stars. There are some rapid pulsators given the designation of hybrid $\gamma$ Dor and $\delta$ Scuti Pulsators \citep{grigahcene2010}, however these types of stars are also observed to be rapid rotators, which is not the case for the $\gamma$ Vir system. \citet{patsourakos19} observed the Sun at 3 mm with ALMA and a $2\,$s cadence. They detected spatially resolved fluctuations in the T$_{B}$ at the few hundred K level on timescales of several minutes. The fluctuations lagged behind $1600\, \rm \AA$ observations by $\sim100$ s and may be due to propagating sound waves in the Solar atmosphere. While observations of stars other than the Sun with this level of time resolution would not be possible due to signal-to-noise constraints, this presents a possible explanation for the variability in $\gamma$ Vir, should it indeed be real.

\subsection{Relevance for debris disk studies}

An unconstrained submm-cm stellar spectrum can lead to an inaccurate interpretation of the presence of debris. This is clearly highlighted if we consider the $\gamma$ Vir system. There are three sources present in $\gamma$ Vir (the two stars and a propsoed background galaxy) that would indeed contribute to uncertainty in the flux from a given component for a lower resolution telescope, but to illustrate this example we will only consider the flux from $\gamma$ Vir B in Band 6. 

If we take the highest measured Band 6 flux value of 356 $\mu$Jy, and assume that there is an underlying debris belt, we can use other standard estimates of the stellar emission to find out how much ``excess" there is. For example, if we assume a Sirius-like emission profile \citep[e.g.,][]{white19}, characterized by $\sim0.6\,\rm T_{B}$, or a simple full blackbody extrapolation of T$_{B}$ of from the optical photosphere temperature (i.e., $100\%$ $T_{B}$), we would get an excess of 180 $\mu$Jy and 65 $\mu$Jy, respectively. If we assume this emission comes from an asteroid belt-type structure, then we can get rough estimate of the total amount of debris. This asteroid belt would be unresolved (i.e., located within the synthesized beam) so we can say it is at a radii of $<9$ au. Adopting the methods in \cite{white_fom}, a lower level total debris mass can be calculated by assuming the excess emission is only coming from $\sim$mm grains located in a single ring at (along with some other basic idealized assumptions such as as the grains being perfect emitters and a density of $2.5\,\rm g\,cm^{-3}$). For simplicity, we assume the asteroid belt is located at 4 au, or half the resolvable radii. For a $\gamma$ Vir-like star with excesses listed above we would expect $2-5\times10^{-5}\,\rm M_{\oplus}$ of mm debris. If this ``excess" were indeed coming from a debris disk, then there should in theory be a full size distribution of grains up to asteroid-sized objects. Again adopting the methods in \cite{white_fom}, we can assume the disk is spread out between $2-4$ au, is populated by grains ranging from $10\,\mu$m to 25 km with a size distribution of $q\sim3.5$, and the emission/absorption efficiency of the grains is dependent on the grain's size. This would lead us to infer $0.02-0.07\,\rm M_{\oplus}$ of total debris in the disk, making it $\sim100\times$ larger than the asteroid belt in our Solar system.

While the above calculations do indeed make some generous and far reaching assumptions given the lack of information, none of them are out of context for estimating the total mass of a debris disk. If the observed flux at $\sim$mm wavelengths was, for example, $10-100\times$ larger than the expected stellar emission for a given system, then the above approach would likely be used to better characterize the debris. This scenario highlights that it is imperative to have an accurate stellar emission model when studying unresolved circumstellar debris. This is particularly true if currently oversubscribed facilitates, such as ALMA, and future facilities such as the \textit{next generation Very Large Array} (ngVLA) are to be used to their full potential in studying unresolved debris structures \citep{white_ngvla}.

\section{Summary}\label{sec:sum}

In this paper, we presented ALMA observations of $\gamma$ Lep, $\gamma$ Vir A, and $\gamma$ Vir B at 0.87 and 1.29 mm. We used the data to inform a KINICH-PAKAL stellar atmosphere code to provide the first semi-empirical models of F-type stars constrained by mm data. The models show a brightness temperature minimum in the sub-mm and a sharp rise at wavelengths longer than $\sim1\,\rm mm$. While we cannot directly confirm or reject the presence of short-term variability within the $\gamma$ Vir system, variability appears to be present in one or both of the stars. These stellar spectra also highlight the potential bias on debris disk studies and demonstrate how unconstrained stellar emission can lead to the inference of circumstellar debris. A comprehensive catalog of stellar spectra that covers submm-cm wavelengths is necessary to fully understand stellar atmospheric processes and accurately study unresolved debris features.

\acknowledgments

We thank the anonymous referee for the feedback on this manuscript. We thank Ryley Hill for useful discussion on the submillimeter behavior of extragalactic sources. JAW acknowledges support from the European Research Council (ERC) under the European Union's Horizon 2020 research and innovation program under grant agreement No 716155 (SACCRED).  FT acknowledges Conacyt-254497 (Ciencia B\'asica) and the Center of Supercomputing of the National Laboratory of Space Weather in M\'exico. AGH and ACB acknowledge support from an NSERC DG and the CRC program. AMH is supported by a Cottrell Scholar Award from the Research Corporation for Science Advancement, This paper makes use of the following ALMA data: ADS/JAO.ALMA\#2017.1.00698.S and ADS/JAO.ALMA\#2018.1.01149.S.  ALMA is a partnership of ESO (representing its member states), NSF (USA) and NINS (Japan), together with NRC (Canada), MOST and ASIAA (Taiwan), and KASI (Republic of Korea), in cooperation with the Republic of Chile. The Joint ALMA Observatory is operated by ESO, AUI/NRAO and NAOJ. The National Radio Astronomy Observatory is a facility of the National Science Foundation operated under cooperative agreement by Associated Universities, Inc.

\facilities{ALMA}


\software{{\scriptsize CASA 5.4.1} \citep{casa_reference} 
          }

\appendix

\begin{figure}
\centering
\includegraphics[width=\textwidth]{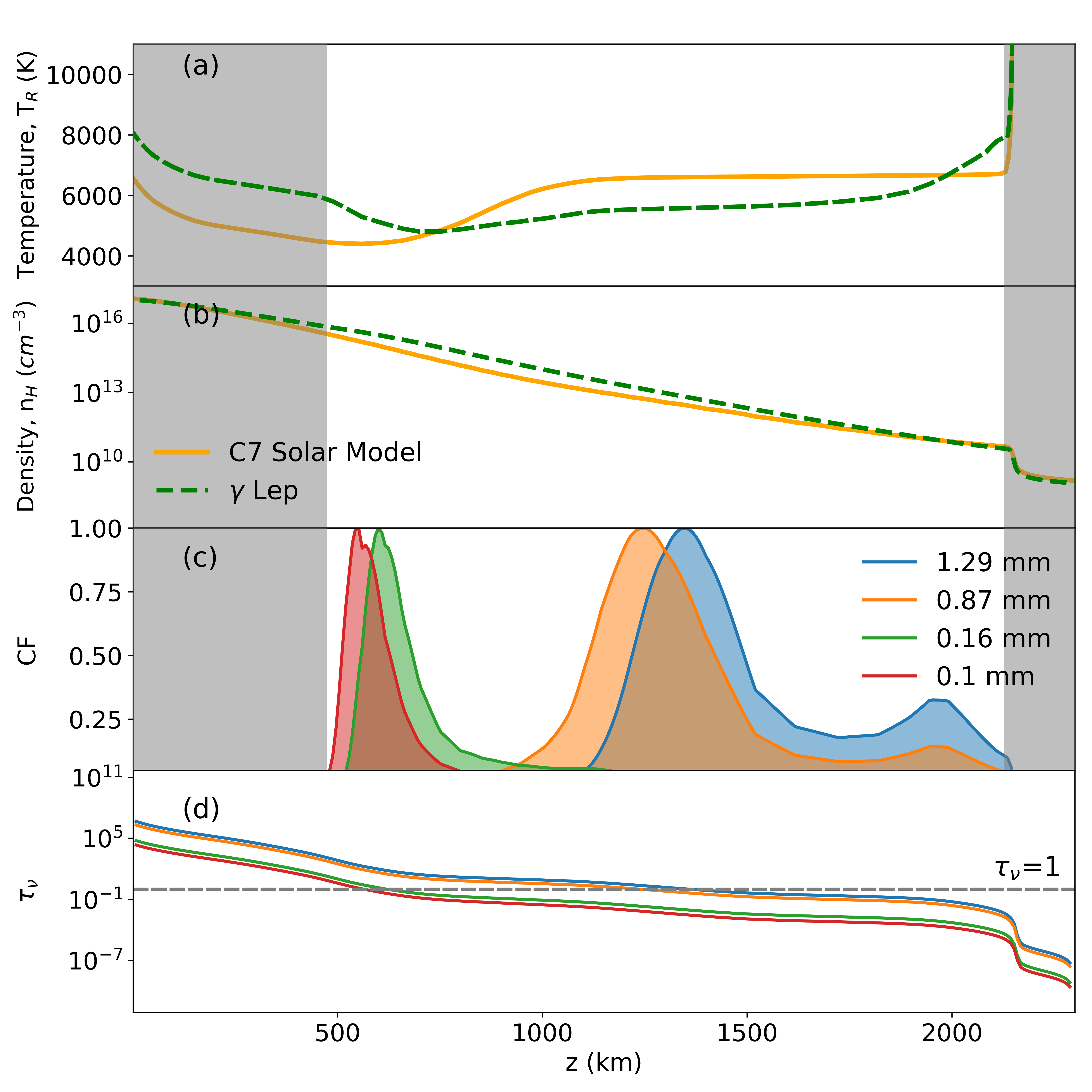}
\caption{$\gamma$ Lep model. 
\emph{(a)}: The green dashed line shows the $\gamma$ Lep temperature profile obtained with KINICH-PAKAL. The orange solid line is the Solar average semi-empirical model C7 \citep{avrett}.  
\emph{(b)}: Density profile for $\gamma$ Lep (green dashed line) and comparison with average Solar values (solid orange line).
\emph{(c)}: Normalized CF for $\gamma$ Lep model for the observable wavelengths. For 1.29 mm the maximum contribution occurs at 1344 km over the photosphere and present a second peak around 1968 km with a contribution of 33\%. At 0.87 mm the CF has the maximum at 1248 km and a second peak at 1968 km with a contribution of 15\%. For 0.16 mm and 0.10 mm, the CF presents one peak at 600 km and 544 km respectively.
\emph{(d)}: Optical depth at 0.19 mm (red line), 0.16 mm (green line), 0.87 mm (orange line), and 1.29 mm (blue line).
\label{GLP}
}
\end{figure}

\begin{figure}
\centering
\includegraphics[width=\textwidth]{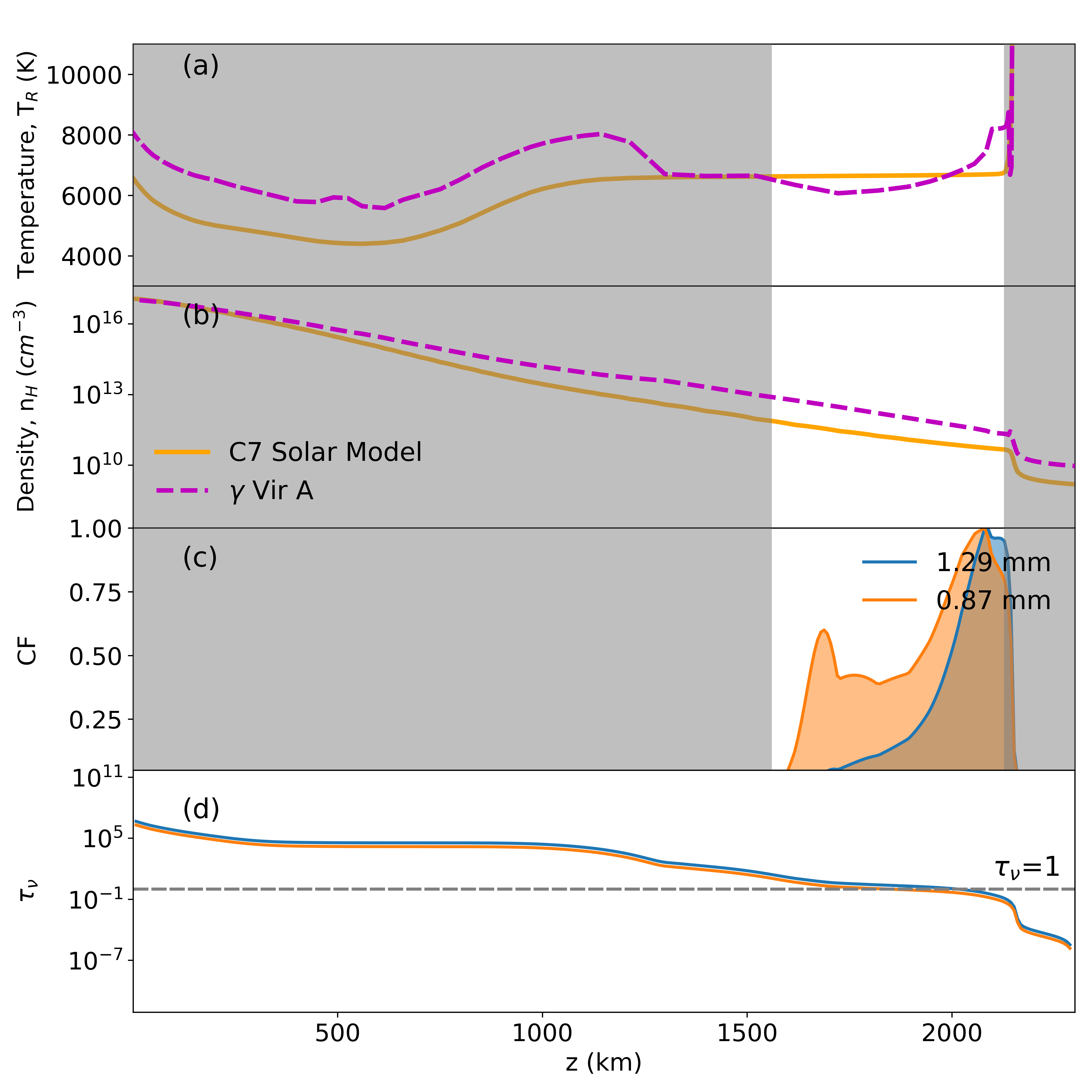}
\caption{$\gamma$ Vir A model.
\emph{(a)}: The magenta dashed line shows the $\gamma$ Vir A temperature profile obtained with KINICH-PAKAL. The orange solid line is the Solar model.  
\emph{(b)}: Density profile for $\gamma$Vir A (magenta dashed line) and comparison with average Solar values (solid orange line).
\emph{(c)}: Normalized CF for $\gamma$ Vir A for the observable wavelengths. At 1.29 mm the maximum contribution occurs at 2080 km over the photosphere. At 0.87 mm, the CF has a maximum at 2080 km with a second smaller peak at 1688 km with a contribution of 60\%.
\emph{(d)}: Optical depth at 0.87 mm (orange line) and 1.29 mm (blue line).
\label{GVA}
}
\end{figure}

\begin{figure}
\centering
\includegraphics[width=\textwidth]{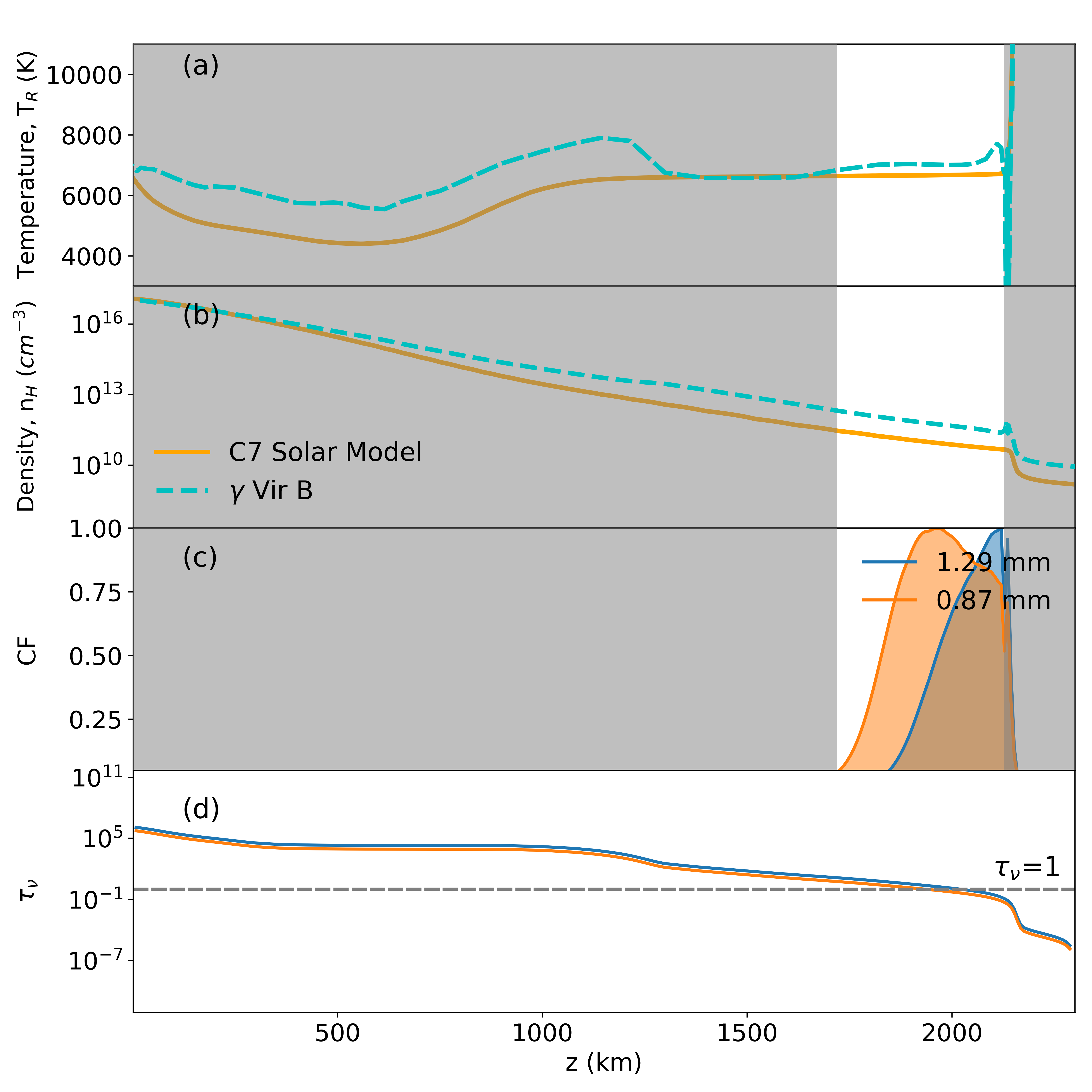}
\caption{ $\gamma$ Vir B model. 
\emph{(a)}: The cyan dashed line shows the $\gamma$ Vir B temperature profile obtained with KINICH-PAKAL. The orange solid line is the Solar model.  
\emph{(b)}: Density profile for $\gamma$ Vir B (cyan dashed line) and comparison with average Solar values (solid orange line).
\emph{(c)}: Normalized CF for $\gamma$ Vir B for the observable wavelengths. At 1.29 mm, the maximum contribution occurs at 2120 km over the photosphere. At 0.87 mm, the CF has a maximum at 1968 km. Both wavelengths just have one peak.
\emph{(d)}: Optical depth at 0.87 mm (orange line) and 1.29 mm (blue line).
\label{GVB}
}
\end{figure}

In Figs.\,\ref{GLP}, \ref{GVA}, and \ref{GVB} we show additional modelling results obtain with the KINICH-PAKAL framework. In each figure, we show the plasma temperature (T$_{R}$), hydrogen density (n$_{H}$), contributions functions (CF), and opacity ($\tau_{\nu}$) as a function of height in the chromosphere. In all of the plots, we include the Solar average semi-empirical model \citep[model C7 from][]{avrett} which were used as initial conditions for each star. 

\end{document}